\documentclass[amsmath,preprintnumbers,10pt,twocolumn,prl]{revtex4-1}
\usepackage{amsbsy}
\usepackage{graphicx}
\usepackage{color}
\usepackage{subfigure}

\usepackage{verbatim}

\begin{document}
\title{Design principles for non-equilibrium self-assembly}
\author{Michael Nguyen and Suriyanarayanan Vaikuntanathan}
\affiliation{Department of Chemistry and The James Franck Institute, The University of Chicago, Chicago, IL, 60637}

\begin{abstract}

We consider an important class of self-assembly problems and using the formalism of stochastic thermodynamics, we derive a set of design principles for growing controlled assemblies far from equilibrium. The design principles constrain the set of structures that can be obtained under non-equilibrium conditions. Our central result provides intuition for how equilibrium self-assembly landscapes are modified under finite non-equilibrium drive.  
\end{abstract}

\maketitle

%Uncovering the principles of self-assembly in far from equilibrium conditions remains an important open question in statistical mechanics. 

The fields of colloidal and nanoscale self-assembly have seen dramatic progress in the last few years. Indeed experimental and theoretical work has elucidated design principles for the assembly of complex three dimensional structures~\cite{Ke2012,Jones2015,Reinhardt2014,Hedges2014}. Most of these advances, however, are based on an equilibrium thermodynamic framework: the target structure minimizes a thermodynamic free energy~\cite{Hormoz2011}. 
Understanding the principles governing self-assembly and organization in far from equilibrium systems remains one of the central challenges of non-equilibrium statistical mechanics~\cite{Weber2012,Mann2009,Whitelam2014a,Sanz2007,Rabani2003,Ursell2014,Levandovsky2009,Hagan2006}. 
In this letter, we show that design principles can be derived for a broad class of non-equilibrium driven self-assembly processes. Our central result constrains the set of possible structures that can be achieved under a non-equilibrium drive.

Imagine a self assembly process in which interactions amongst the various monomers are described by a set of energies $E^{\rm eq}$. The ratio of association and dissociation rates is set by a combination of interaction energies and chemical potentials $\{\dots \mu_i\dots \}$ of the monomers. This generic setup is sufficient to describe many self assembly processes. Examples include: growth of crystals from solution by nucleation~\cite{Sanz2007}, growth dynamics of cell walls~\cite{Jiang2011}, growth of multicomponent assemblies~\cite{Hedges2014} and growth dynamics of biological polymers and filaments~\cite{Andrieux2008}. 
The chemical potential controls the growth of the assembly. If the chemical potential is tuned to a \textit{coexistence} value such that the assembly grows at an infinitesimally slow rate, then the structure of the assembly and its composition can be predicted  by computing the equilibrium partition function and free energy $G^{\rm eq}$ appropriate to the set of interaction energies. 

For values of the chemical potentials more favorable than the coexistence chemical potential, the assembly grows at a non-zero rate. In such instances, the structure and composition of the growing assembly might not have sufficient time to relax to values characteristic of the equilibrium partition function~\cite{Sanz2007,Kim2009,Auer2001}. Defects are accumulated as the self assembled structure grows at a non-zero rate. The time taken for a defect to anneal increases rapidly with distance from the interface of the growing structure. Due to the resulting kinetically-trapped states the crystal can assume structures very different from those representative of the equilibrium state~\cite{Sanz2007,Kim2009,Auer2001}.

By applying the second law of thermodynamics and the formalism of stochastic thermodynamics, we derive a surprising thermodynamic relation that is applicable to the above mentioned kinetic processes. This relation provides constraints on the configurations that are achievable in a non-equilibrium self assembly process,
\begin{figure}[tbp]
\centering
\includegraphics[width=1\linewidth,angle=0]{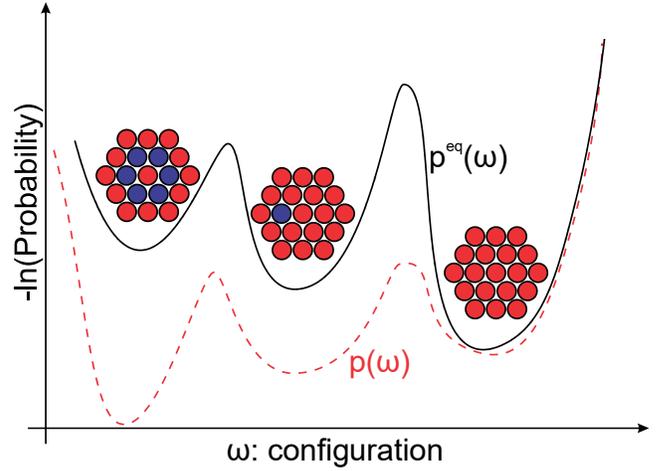}
\caption{ Schematic of the self-assembly problem considered in the paper. The probability of observing structural and compositional fluctuations in the growing assembly, $p(\omega)$ can be different from the canonical distribution $p^{\rm eq}(\omega)$ specified by the interaction energies $E^{\rm eq}$. Our central result, Eq.~\ref{eq:central} constrains the set of compositional fluctuations that can be achieved under a chemical potential drive $\delta \mu$.}
\label{fig:one}
\end{figure}

\begin{equation}
\label{eq:central}
\frac{d \langle N \rangle_t}{dt }\left[\delta \mu- \frac{D[p_N(\omega)||p^{\rm eq}_N(\omega)]}{N}\right]\geq 0 \,,
\end{equation}
where $N$ is the size of the assembly at some instant of time, $d \langle N \rangle_t /dt $ is the average rate of the growth of the assembly, $p_N(\omega)$ is probability distribution associated with a configuration $\omega$ in the growing assembly, $p^{\rm eq}_N(\omega)\equiv \exp[-(E^{\rm eq}(\omega)-G_N^{\rm eq})/k_BT]$ is the equilibrium probability distribution obtained when the assembly is grown at an infinitesimally slow rate and $D[p||q]=\int p \ln p/q \geq 0$ is the relative entropy between distributions $p$ and $q$. The relative entropy is a measure of distinguishability between distributions $p$ and $q$. It is zero only when the two distributions are identical and is nonzero otherwise. We have assumed that the chemical potential of each monomer is the same and exceeds the equilibrium coexistence value by $\delta \mu$ (i.e. the concentrations of monomers in solution exceeds the equilibrium concentration required for assembly). The chemical potential difference $\delta \mu$ provides the non-equilibrium driving force for self assembly. 

An alternative and presumably more practical formulation of the central result is in terms of interaction energies. Imagine that we wish to generate compositions and structures in the growing assembly which are characteristic of a Hamiltonian $E^{\rm eff}$ different from the Hamiltonian governing the interactions between species $E^{\rm eq}$. The central result places a bound on the minimum required excess chemical potential $\delta \mu$ required to achieve such an assembly,
\begin{equation}
\frac{d\langle N \rangle_t}{dt}   \left[ \frac{-G_N + G_N^{\rm eq}+ \langle E^{\rm eff}-E^{\rm eq} \rangle_N}{N} +\delta \mu \right]  \geq 0\,.
\label{eq:central2}
\end{equation}
Here $G_N$ is the free energy of $N$ particle system described by the Hamiltonian $E^{\rm eff}$. The free energies differences can either be computed directly from simulations or estimated using an analytical framework.

Hence, given a chemical potential drive Eq.~\ref{eq:central}, Eq.~\ref{eq:central2} constrain the set of allowed non-equilibrium structures found in the assembly. Alternately, given a target distribution $p_N(\omega)$ or a target effective Hamiltonian $E^{\rm eff}$, the central result sets a bound on the minimum chemical potential driving force required to achieve the assembly. 
If the chemical potentials of the monomers can be varied, $\delta \mu$ in Eq.~\ref{eq:central} and Eq.~\ref{eq:central2} is replaced by an average, $\langle \sum_i\delta \mu_i \rangle_N/N$. The  bound can be used to variationally optimize non-equilibrium driving forces $\{\dots \mu_i\dots\}$ that result in configurations characteristic of a desired effective energy landscape.

\section{Thermodynamic bounds for non-equilibrium self assembly}

We now outline the derivation of the central result. This derivation is based on a generalization of the framework introduced in Ref.~\cite{Andrieux2008}.  
We begin by invoking one of the central results of stochastic thermodynamics~\cite{Esposito2010,Andrieux2008,Schnakenberg1976} for a system in contact with a heat bath,
\begin{equation}
\label{eq:StochasticThermo}
\frac{d S_{\rm tot}}{d t}\equiv \frac{d S}{d t} +\frac{d S_e}{dt }   \geq 0
\end{equation}
where $S$ is the entropy of the system and $S_e$ is the entropy due to energy exchange between the system and environment. Eq.~\ref{eq:StochasticThermo} is simply a statement of the second law of thermodynamics. 

The growing self assembling structure will be the thermodynamic system for the purposes of this paper. The system is allowed to exchange material and energy with the environment. Let $P_t(\omega,N)$ denote the probability distribution associated with the observing a system size $N$ and a microscopic configuration $\omega$ at a particular instant of time. The entropy of the system $S$ is given by 
\begin{equation}
S=-k_B \sum_{\omega,N} P_t(\omega,N) \ln P_t(\omega,N) 
\end{equation}
where $k_{B}$ is the Boltzmann constant. 

To proceed, we decompose the distribution $P_t(\omega,N)$ as $P_t(\omega,N)\equiv P_t(N)p_N(\omega)$ where $P_t(N)$ and $p_N(\omega)$ are both normalized probability distributions. In performing the decomposition we have assumed that the system has reached a steady state and that distribution of compositional fluctuations for a given system size, and that $p_N(\omega)$ is independent of the time $t$. This decomposition is the main assumption behind our theoretical approach. 
With this assumption, we can associate an effective energy function $E^{\rm eff}(\omega)$ with the distribution $p_N(\omega)$,  $E^{\rm eff}(\omega)\sim -\ln  p_N(\omega)$. This is simply a statistical defining relation for  the energy function (analogous to a potential of mean force). The energy function $E^{\rm eff}(\omega)$ does not control the dynamics of the system.  This effective energy function can have a form very different from that of the interactions specified by the interaction Hamiltonian $E^{\rm eq}(\omega)$. 
 
\begin{figure*}[tbp]
\centering
\includegraphics[width=1\linewidth,angle=0]{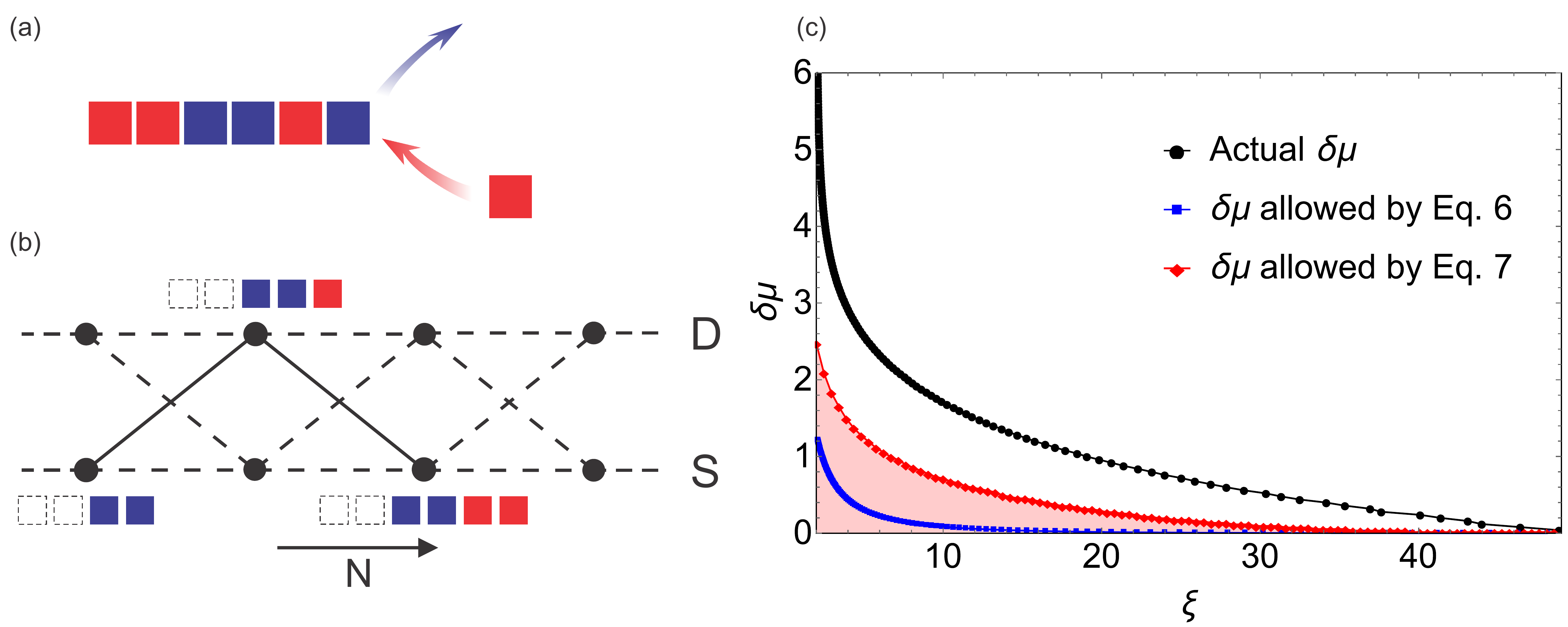}
\caption{Application of the bounds to 1D fiber growth model (a) Schematic of the 1D polymer growth process. The polymer is assembled from a bath composed of two monomer types (Red and Blue). (b) Schematic of the effective Markov state model. This Markov state model resolves the  nature of terminal bond in the self-assembled system (vertical rungs), $S\equiv$ bond between like monomers and $D\equiv$ bond between unlike monomers, and the number of particles in the self assembled system (horizontal axis). The rates of transitions between the mesoscopic states resolved in this effective model depend on the composition of the assembly and are described in the (SI). We demonstrate that this effective model captures the dynamics and compositional fluctuations of the self assembly process. (c) Comparison between the lower bounds of $\delta\mu$ obtained from Eq.~\ref{eq:centralresult1} and Eq.~\ref{eq:centralresult2} and the value of $\delta\mu$ obtained from simulations. The equilibrium domain length, $\xi_0$, for this plot is 50. The red shaded region represents the region that is disallowed by Eq.~\ref{eq:centralresult2}. Our thermodynamic bounds are valid and inspite of their minimal nature, do reasonable job of predicting the compositional fluctuations in the assembly.}
\label{fig:Xi}
\end{figure*}

Using this decomposition and the total average entropy production rate in a self assembly process, $d S_{\rm tot}/dt $ can be written as
\begin{equation}
\label{eq:define}
\begin{split}
\frac{d S_{\rm tot}}{d t} &=\frac{d\langle N \rangle_t}{dt}   \left[\delta \mu- \frac{G_N - G_N^{\rm eq}+ \langle E^{\rm eq}-E^{\rm eff} \rangle_N}{N} \right]\\ & = \frac{d\langle N \rangle_t}{dt}  \left[\delta \mu- \langle \epsilon_{\rm diss}\rangle\right]\geq 0 \,.
\end{split}
\end{equation}
A closer investigation reveals that $ \langle \epsilon_{\rm diss}\rangle\geq 0$\footnote{The proof for this statement is simple once we realize that $\langle \exp [-(E^{\rm eq} -E^{\rm eff})]\rangle_N =\exp(-(G_N^{\rm eq}-G_N))$. Using Jensen's inequality, we obtain the bound on  $\langle \epsilon_{\rm diss}\rangle$}.

Hence, in order to have a growing assembly with a distribution of compositional fluctuations characteristic of the energy function $E^{\rm eff}$, the chemical potential difference $\delta \mu$ must at least exceed $\langle \epsilon_{\rm diss}\rangle$, 
\begin{equation}
\label{eq:centralresult1}
\delta \mu \geq \langle \epsilon_{\rm diss}\rangle\,.
\end{equation}
A simple algebraic manipulation allows us to rewrite Eq.~\ref{eq:centralresult1} as Eq.~\ref{eq:central}. When the chemical potentials of the monomers contributing to the assembly are unequal, $\delta \mu$ is replaced by the average excess chemical potential for the growing assembly, $\langle \sum_i\delta \mu_i \rangle_N/N$.

The structure of Eq.~\ref{eq:centralresult1} is reminiscent of the variational Gibbs-Bogoliubov-Feynman inequality generalized for non-equilibrium systems. As such it can be applied to obtain bounds on optimal non-equilibrium driving forces that can maximize the yield of desired structures. Finally, our central results show how the average entropy production rate in a self assembly process is coupled to the compositional fluctuations in the self assembled system. Recent work by Barato, Seifert and coworkers  has demonstrated that the entropy production rate bounds the fluctuations of various dynamical currents generated in  a non-equilibrium processes~\cite{Barato2015,Barato2016,Gingrich2016}. In the next section, using a model system, we demonstrate that the bound imposed by the entropy production rate on cumulants of fluctuations of the growth rate of the assembly lead to a set of tighter design principles,
\begin{equation}
\label{eq:centralresult2}
\delta \mu -\langle  \epsilon_{\rm diss} \rangle \geq \frac{2\langle \dot{N} \rangle}{t \langle (\delta {\dot N})^2 \rangle}\,,
\end{equation}
where $\langle {\dot N }\rangle\equiv N/t$ denotes the average rate of growth of the assembly, $\langle (\delta {\dot N})^2 \rangle$ denotes the variance in the growth rate of the assembly.

We will now illustrate the result on two space filling lattice based model systems. In each example, the distribution of compositional fluctuations in the growing assembly is modified when the assembly is grown at a non-zero rate. Our bound acts as a design principle and provides intuition into how the equilibrium landscape of compositional fluctuations is modified under a non-equilibrium drive. We emphasize that while these examples focus on distributions of compositional fluctuations, our central results are valid more generally and also apply to statistics of structural fluctuations~\cite{Sanz2007}. 
We choose the lattice based models as examples because they are analytically tractable and serve to illustrate the utility of Eq.~\ref{eq:centralresult1} and Eq.~\ref{eq:centralresult2}.   
%VarJ.pdf
\section{Exactly solved model of non-equilibrium polymer assembly}
As our first example, we will consider a model of one-dimensional polymer assembly. For simplicity, the system that we are studying will contains two types of monomers with equal concentrations. In addition, only nearest neighbors interaction is allowed between monomers. In scenarios where the polymer is nucleated from a solution, the concentration of monomers or their chemical potential provides the non-equilibrium driving force for the assembly. If the average length of monomer domains in a polymer at equilibrium is $\xi_0$, our results impose a constraint on the excess chemical potential required to achieve assemblies in which the average length of monomer domains is $\xi$ (SI),

\begin{equation}
\delta \mu \geq \left[(1-\frac{1}{\xi} )\ln \frac{\xi-1}{\xi_0-1} + \ln \frac{\xi_0}{\xi}\right]\equiv \langle\epsilon_{\rm diss}\rangle 
\label{eq:centralisingone}
\end{equation}
where $\delta \mu$ is the excess chemical potential and we have assumed that the two monomers have the same chemical potential. The compositions in such an assembly differ from those obtained close to equilibrium ($\delta \mu \sim 0$) when the polymer grows at an infinitesimally slow rate. Our thermodynamic bound doesn't depend on the kinetic model chosen. Further, as discussed previously, in cases with unequal chemical potentials, Eq.~\ref{eq:centralisingone} can be generalized by simply replacing $\delta \mu$ with the average excess chemical potential.

We theoretically verify the bound for the polymer assembly  problem by mapping the dynamics of the self assembly process onto an effective Markov state model (Fig \ref{fig:Xi} (b)). This effective model resolves the terminal bond in the self-assembled system (vertical rungs) and the number of particles in the self assembled system (horizontal axis). The details of the various transition rates between the mesoscopic states are provide in the supplemental information (SI). In the SI, we also demonstrate that the Markov state caricature accurately describes the dynamics and compositional fluctuations of the self assembly process. Further, we demonstrate that the expression for the entropy production rate, $d S_{\rm tot}/d t$ computed from this Markov state caricature indeed equals 
\begin{equation}
\frac{d S_{\rm tot}}{d t} =\frac{ d N}{dt}(\delta\mu -\langle\epsilon_{\rm diss}\rangle\,)
\end{equation} 
where $\langle\epsilon_{\rm diss}\rangle$ is defined in Eq.~\ref{eq:centralisingone}. This calculation verifies the thermodynamic bound and its underlying assumptions for the one dimensional polymer assembly model.

The Markov state caricature allows us to map the self assembly process onto the dynamics of a random walker on a graph. In the SI, we verify that the effective mean field Markov state caricature manages to reasonably describe even the second cumulant of fluctuations in the growth rate. Barato, Seifert and coworkers have recently demonstrated~\cite{Barato2015,Barato2016,Gingrich2016} that fluctuations of various dynamical currents in such Markov state process are bound by the entropy production rate. As demonstrated in the SI, we adapt these bounds to show that 
\begin{equation}
\label{eq:centrallinear}
\delta \mu \geq \left[(1-\frac{1}{\xi} )\ln \frac{\xi-1}{\xi_0-1} + \ln \frac{\xi_0}{\xi}\right] + \frac{2\langle {\dot N} \rangle}{t \langle (\delta {\dot N})^2 \rangle}
\end{equation}
where $\langle \dot N \rangle$ is the average growth rate of the polymer and $\langle (\delta \dot N)^2 \rangle$ is the variance of the growth rate.  

In Fig.~\ref{fig:Xi}(c), we demonstrate the two thermodynamic bounds in Eq.~\ref{eq:centralisingone}, Eq.~\ref{eq:centrallinear} for a particular kinetic model of polymer growth~\cite{Whitelam2012}. 
Including information from higher order cumulants does indeed strengthen the bound. These demonstrations prove the theoretical validity and effectiveness of our bound and pave the way for applications to more complex systems.

\begin{figure*}[tbp]
\centering
\includegraphics[width=1.0\linewidth,angle=0]{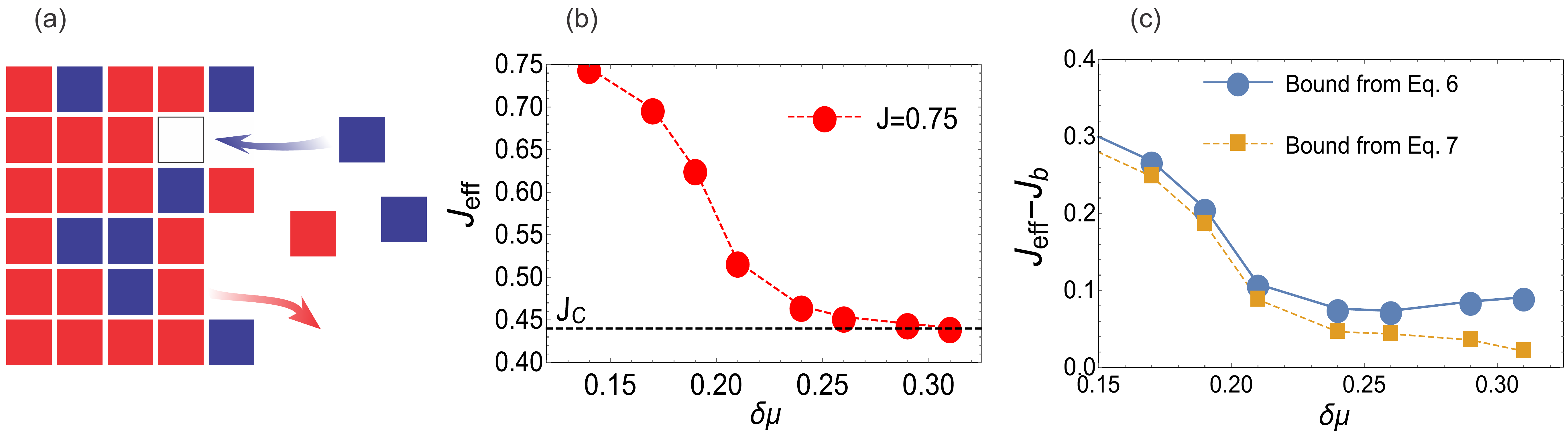}
\caption{ Application of the bounds to the non-equilibrium two dimensional growth process (a) We assemble a two dimensional structure from a monomer bath containing red and blue blocks. The energy of interaction between similar monomers is $\epsilon_s$ while that between dissimilar monomers is $\epsilon_d$. As discussed in the text, when the assembly is grown at equilibrium, the statistics of compositional fluctuations in the assembly are equivalent to that of an Ising magnet with coupling constant $J=\frac{\epsilon_s-\epsilon_d}{2}$. (b)  For assemblies grown away from equilibrium, we analyzed the compositional fluctuations and using the analytical solution for a two dimensional Ising magnet, compute the effective coupling constant $J_{\rm eff}$ as a function of the excess chemical potential $\delta\mu$. The equilibrium coupling constant was set to $J=0.75$. As the system deviates more from equilibrium by increasing $\delta\mu$, the $J_{eff}$ decreases and reaches a critical value $J_c$ for $\delta\mu \geq 0.30$ . (c) We computed bounds for the effective coupling constant, $J_b$ using Eq.~\ref{eq:centralresult1} and Eq.~\ref{eq:centralresult2} at J = 0.75. As evidenced by the difference $J_{\rm eff}-J_b$, the bounds are able to offer reasonable intuition for the statistics of compositional fluctuations in the growing non-equilibrium assembly. The bounds work particularly well for values of excess chemical potential $\delta\mu$ close to the critical value $\delta \mu_c\approx 0.30$. This result is striking given the minimal nature of our model.}
\label{fig:Ising2Deffective}
\end{figure*}

Specifically, for systems with more than two monomer types or for monomers with longer ranged interactions, it is more convenient to express $\langle \epsilon_{\rm diss} \rangle$ in terms of the actual $E_{\rm eq}$ and effective $E_{\rm eff}$ interaction Hamiltonians.  While effective Markov state caricatures for such systems can be complex and analytically intractable, our thermodynamic bounds are easily accessible and provide very useful intuition for the compositional fluctuations of the non-equilibrium assembly. We now demonstrate these ideas by considering a two-dimensional analogue of the fiber growth process described above~\cite{Whitelam2014} (Fig.~\ref{fig:Ising2Deffective} (a)).

\section{Non-equilibrium two dimensional self assembly process}

As a natural generalization of the one-dimensional polymer model, we imagine a two-dimensional assembly growing at one end as described in Fig \ref{fig:Ising2Deffective} (a). We again imagine two monomer types with interactions between like monomers set by the energy scale $\epsilon_s$ and interactions between unlike monomers set by the energy scale $\epsilon_d$. The details of the kinetic model are described in Methods. 
 When the chemical potential is set to the intensive equilibrium free energy of the assembly, $\mu_{\rm coex}\approx 2\epsilon_s-k_{\mathrm{B}}T \ln 2 $, the assembly doesn't grow on average. Assigning spin $s=1$ to the red monomers and spin $s=-1$ to the blue monomers, the statistics of the compositional fluctuations in the equilibrium assembly are equivalent to that of an Ising model with coupling constant $J=\frac{\epsilon_s-\epsilon_d}{2}$. We will work in regimes where the coupling constant is above the critical coupling constant of the Ising model, $J>J_c\approx 1/2.26$. We are interested in the statistics of compositional fluctuations as the growth rate of the assembly is tuned by changing the chemical potential. 
 
We begin by assuming that the compositional fluctuations in the growing assembly can be described by an effective nearest neighbor coupling constant $J_{\rm eff}$. We performed simulations in which we extracted the value of $J_{\rm eff}$ as a function of the excess chemical potential $\delta \mu$ by analyzing compositional fluctuations in the growing assembly (Methods). The values of $J_{\rm eff}$ obtained from simulations are plotted in Fig.~\ref{fig:Ising2Deffective}(b). The underlying Ising model had a coupling constant $J=0.75$. When the excess chemical potential is close to $\delta\mu\approx 0.30$, the value of the effective coupling constant computed from simulations reaches the critical value of the two dimensional Ising model, $J_c$. 

Using Eq.~\ref{eq:centralresult1} and Eq.~\ref{eq:centralresult2} we analytically computed a lower bound $J_b$ on the values of the effective coupling constant as a function of $\delta \mu$. In Fig.~\ref{fig:Ising2Deffective} (b), we plot the difference $J_{\rm eff}-J_b$ for various values of the excess chemical potential. The bound obtained from Eq.~\ref{eq:centralresult2} does a reasonable job of predicting the effective coupling constant. Including information about higher order cumulants again improves the bound and provides a close to accurate estimate of the effective coupling constant $J_{\rm eff}$. In particular, the bound obtained by including information form the higher cumulants (Eq.~\ref{eq:centralresult2}) does particularly well for values of $\delta \mu$ around the critical value, $\delta \mu_c\approx0.30$. This close agreement between the bound predicted by theory and the effective value computed from simulations in the critical regime is surprising given the minimal nature of our theory. 

We found similar results for other values of the coupling constant $J$. Strikingly, for all the regimes considered, the bounds are particularly accurate when the assembly is close to criticality. The results illustrate the utility of our central relations as design principles for non-equilibrium self assembly.

\section{Applications and Conclusion}

%. However, we lack general  

Using the framework of stochastic thermodynamics and physically justified approximations, we have put forward a general predictive framework for non-equilibrium self assembly. Our central results, Eq.~\ref{eq:centralresult1} and Eq.~\ref{eq:centralresult2} are built around a simple and intuitive expression for entropy production in non-equilibrium self assembly process. This expression relates compositional fluctuations in the self assembled system to the chemical potential excess driving assembly process. Given a set of programmed interactions between monomers and structures specified by the energy function $E^{\rm eq}$, our central result constrains the set of allowed structures that can be obtained under a non-equilibrium drive. A unique feature of our result is its thermodynamic character: the bound doesn't depend on the details of the kinetic model chosen. Many recent theoretical and experimental studies have postulated that non-equilibrium protocols can potentially be used to overcome bottlenecks and enhance the yield of desired self assembled structures~\cite{England2015}. Our framework is ideally suited to verify these postulates and explore connections between energy consumption and improved yield of target structures.

We anticipate that our results will find important applications in studies of crystal nucleation~\cite{Auer2001}, self assembly of metal organic frameworks~\cite{Sue2015}, non-equilibrium roughening transitions, and the synthesis of nano crystals using DNA and other biopolymers as building materials. In all these instances, the assembly process can occur far from equilibrium depending on the effective chemical potentials imposed. 
Our result provides a bound on the chemical potentials under which the desired structures can be robustly obtained even far from equilibrium. It also provides crucial intuition into how the equilibrium landscape~\cite{Hormoz2011,Reinhardt2014} and the capacity of the system to store structures~\cite{Murugan2015} is modified under non-equilibrium conditions. 

 Finally, the variational structure of our central results provides a framework to optimally tune the chemical potential driving forces so as to maximize the yield of desired structures. In this context, our results complement recently derived theoretical results that suggest a heteregeous set of excess chemical potentials is crucial for robust nucleation of large complex nanoscale assemblies using DNA origami~\cite{Jacobs2015,Murugan2015a}.

A mismatch of chemical potentials provides the non-equilibrium driving force in Eq.~\ref{eq:centralresult1}. Our results also apply to non-equilibrium error correction mechanisms for biopolymer assembly in which the assembly process is driven by consumption of energy rich ATP/GTP molecules. In such instances, the $\delta \mu$ term is replaced by the appropriate average energy consumption rate. In these biophysical applications, the non-equilibrium driving force generates structures that posses correlations greater than those in equilibrium. While such fibers are not achievable with the kinetic protocols described above, the thermodynamic nature of our bounds allows their application to these biophysical error correction mechanisms. Indeed, Sartori et al~\cite{Sartori2015a} have recently derived a similar bound for a specific model of kinetic proofreading. Our general thermodynamic bounds, particular that in Eq. \ref{eq:centralresult2} can elucidate the tradeoffs between speed accuracy and energy consumption in such non-equilibrium biochemical replication processes.

\section{Methods}
In our 1D polymer growth simulation, we consider a polymer growing in a bath containing two types of monomer. The concentrations of both types are equal to c. The chemical potential $\mu$ is related to the concentration c by the relation: $\mu = -k_BT\ln(c)$. The nearest neighbor monomers are allowed to interact with energy $\epsilon_s$ if they are the same and with energy $\epsilon_d$ if they are different.  We perform kinetic Monte-Carlo simulations in which monomers are added to the polymer with the rate c (or  $\exp(-\beta\mu)$). Monomers are allowed to dissociate from the polymer with the rate $\exp(-\beta\epsilon_i)$ (i here can be either s or d) which depends solely on the composition of the fiber. $\beta$ is set to 1. We use $\phi$ to denote the probability that a particular bond in the assembly is between two like monomers. It is related to the correlation length $\xi$ by: $\phi=1-1/\xi$.

The simulations for the non-equilibrium 2D assembly problem follow the same rules as in the 1D case. However, we used the standard Metropolis Monte-Carlo algorithms for the 2D case.  For the purposes of computing the effective coupling constant $J_{\rm eff}$ we labeled the red particles as spin up $s=1$ and blue particles as spin down $s=-1$. We then computed the average per unit magnetization $m$ in the growing assembly and used the Onsager solution to extract the value of $J_{\rm eff}$. This procedure is only valid for $J_{\rm eff} > J_c$.

%using DNA and other biopolymers as building materials. In this synthesis technique, a set of desired structures can be programmed in by choosing the right set of complimentary DNA strands. These structures represent the ground state of an equilibrium thermodynamic landscape. Depending on the chemical potentials of the monomers, the assembly of the structures can however be a non-equilibrium process. 
\section{Acknowledgements} 

We gratefully acknowledge useful discussions with Dmitri Talapin and support from the University of Chicago.

%\bibliography{citations1}
\bibliographystyle{pnas}

\end{document}